\documentclass[conference]{IEEEtran}
\IEEEoverridecommandlockouts
\usepackage{cite}
\usepackage{amsmath,amssymb,amsfonts}
\usepackage{algorithmic}
\usepackage{graphicx}
\usepackage{textcomp}
\usepackage{color}
\usepackage[hyphens]{url}
\usepackage[stable]{footmisc}
\def\BibTeX{{\rm B\kern-.05em{\sc i\kern-.025em b}\kern-.08em
    T\kern-.1667em\lower.7ex\hbox{E}\kern-.125emX}}
\begin{document}

\title{Estimation of physical activities of people in offices from time-series point-cloud data}

\author{\IEEEauthorblockN{Koki Kizawa, Ryoichi Shinkuma$^{1}$\thanks{$^{1}$Ryoichi Shinkuma is also with Hyper Digital Twins Co., Ltd. (HDT).}, and Gabriele Trovato}
\IEEEauthorblockA{Shibaura Institute of Technology, Japan \\
\{al19121, shinkuma, gabu\}@shibaura-it.ac.jp}
}

\maketitle

\begin{abstract}
This paper proposes an edge computing system that enables estimating physical activities of people in offices from time-series point-cloud data, obtained by using a light-detection-and-ranging (LIDAR) sensor network.
The paper presents that the proposed system successfully constructs the model for estimating the number of typed characters from time-series point-cloud data, through an experiment using real LIDAR sensors.
\end{abstract}


\section{Introduction\footnote[2]{This work was supported in part by NICT (no. 06401). The authors thank LiNew inc. for their advice.}}
\label{sec:introduction}
It is reported that job performance of people in offices depend on their physical activities \cite{9596583}.
Traditionally, normal cameras were straightforward to capture visual information of human actions and poses \cite{8593867}.
LIDAR sensors are also beneficial to three-dimensional (3D) visual information \cite{9795869}.
It is unclear how we estimate physical activities of people in offices from point-cloud data by LIDAR sensors.

This paper proposes an edge computing system that enables estimating physical activities of people in offices by using a LIDAR sensor network.
The proposed system constructs a model for the estimation from a dataset of time-series point cloud data and, using the model, performs the estimation from time-series point cloud data obtained in an office.
%

\section{Proposed system}
\label{sec:proposed}

\subsection{System model}
\label{sec:system}
\begin{figure}[t]
\centerline{\includegraphics[width=80mm]{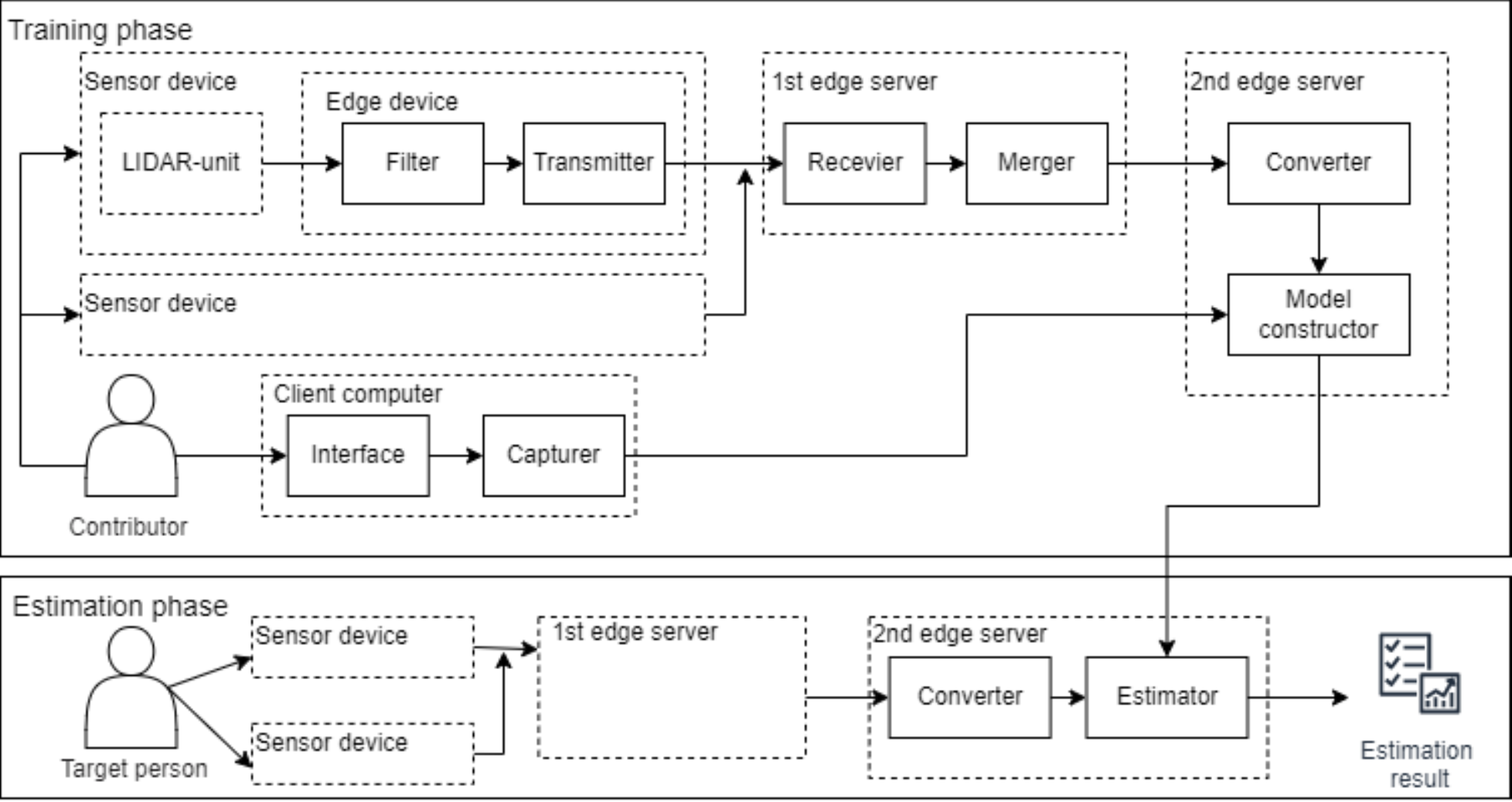}}
\caption{System model.}
\label{systemmodel}
\end{figure}

Fig.~\ref{systemmodel} presents the system model.

The part of the training phase consists of multiple sensor devices, a 1st edge server, a 2nd edge server, and a user interface.
Each sensor device consists of a LIDAR unit and an edge device.
The edge device consists of a filter, and a transmitter, while the 1st edge server consists of a receiver, and a merger as Akiyama et al. presented \cite{9789786}.
The 2nd edge server consists of a converter, and a model constructor.
The user interface consists of an interface and a capturer.
The sensor devices capture time-series point cloud data of the contributor for her or his physical activity.
The filter receives data from the LIDAR unit and removes points in the space not necessary for estimation.
The receiver of the 1st edge server gets data from the transmitter of the edge device.
The merger aggregates data received from multiple edge devices.
The converter processes time-series point-cloud data received from the merger so as to fit the format for the model converter.
Physical activities of a contributor are captured via the interface.
The captured data are also forwarded to the model constructor.
The model constructor builds a model for estimation.

The part of the estimation phase consists of multiple sensor devices, a 1st edge server, and a 2nd edge server.
The sensor devices capture time-series point cloud data of the target person for her or his physical activity.
The estimator uses the model obtained from the model converter for estimation.
It performs the estimation of physical activity from time-series point cloud data of the target person.

\subsection{Methodology}
\label{sec:methodology}
The proposed system constructs a model for estimating physical activities from time-series point-cloud data.
%
%
In this experiment, we consider typing characters via a PC keyboard as a physical activity.
We expect that the model successfully constructed by the proposed system enables estimating the number of typed characters.
However, since point cloud data just represent a set of coordinates of points, to capture the motion of a physical activity we need to introduce a method that associates the point cloud of a frame with the one of the subsequent frames \cite{weng2019baseline}.

In this paper, to simplify computation, we first remove one of the three coordinates (x, y, and z) of each point of point cloud data and project the points to the plane of the other two coordinates.
For instance, if we remove the z-coordinate of points, the points are projected to the x-y plane.
We then convert the two-dimensional point cloud data to a two-dimensional pixel image.
As a result, a sequence of frames of time-series point-cloud data is converted to a sequence of frames of a two-dimensional pixel image, which we call `video frames' hereafter.

Zero-means Normalized Cross-Correlation (ZNCC) has been used to measure the difference between two video frames.
We use ZNCC to capture the motion of video frames \cite{CHEN2020272, NAKHMANI2013315}.
ZNCC is defined as:
\begin{eqnarray}
\label{eq:zncc}
ZNCC=\frac{\sum_{i-1}^{MN} (x_{i}-\bar{x})(y_{i}-\bar{y})}{\sqrt{\sum_{i=1}^{MN} (x_{i}-\bar{x})^2 \sum_{i=1}^{MN} (y_{i}-\bar{y})^2}}, \nonumber
\end{eqnarray}
where $x$ and $y$ are the intensity values of the 1st and 2nd dimensions, and $M$ and $N$ are the number of pixels of 1st and 2nd dimensions.
We obtain ZNCC between $i$-th frame and $i+1$-th frame for the 1st to $I-1$-th frames, where $I$ is the number of frames.
From the distribution of the obtained values of ZNCC, we use  the 90 percentile value as the metric that captures the motion of video frames.

\section{Evaluation}
\begin{figure}[t]
\centerline{\includegraphics[width=1.0\linewidth]{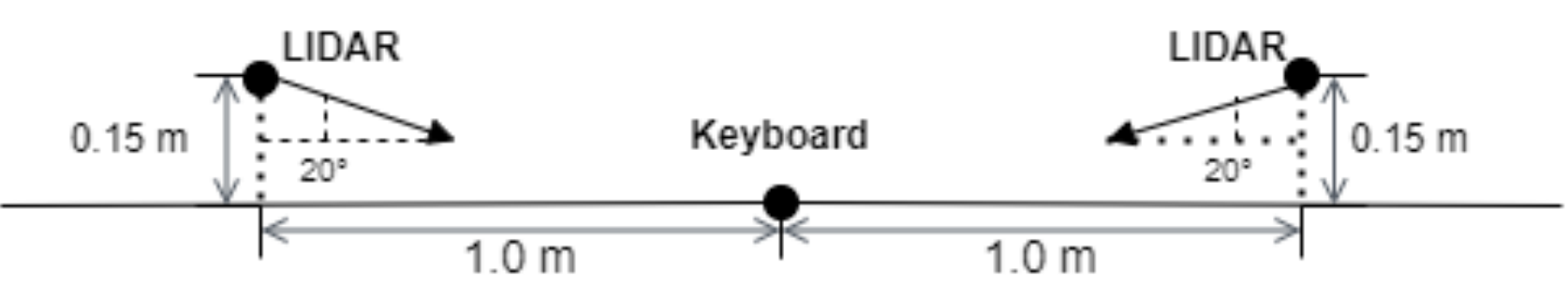}}
\caption{Arrangement of LIDAR sensors for experiment}
\label{fig_ex}
\end{figure}

%
We assume a scenario in which a target person types characters via a PC keyboard in an office.
We prototyped the system presented in Fig.~\ref{systemmodel}.
We used Livox Mid-70 for LIDAR sensors \cite{Mid70}, Jetson Nano for edge devices \cite{Jetsonnano}, and Jetson Xavier NX for edge computers \cite{JetsonNX}.
Fig. \ref{fig_ex} shows the arrangement of LIDAR sensors.
The PC keyboard was located at the center between the two LIDAR sensors.
The distance from the center was 1.0 m.
The LIDAR sensors were tilted to the table with 20 degree.
We used only the data of the (430 mm $\times$ 450 mm $\times$ 170 mm) space right upper of the PC keyboard.
The target person typed characters along the ``myTyping'' question text \cite{myTyping}.
The duration of the experiments was 60 s.
For each experiment, the number of typed characters was counted at the end of the experiment.
Four subjects were volunteers being target persons (subjects A to D).
We obtained 39, 5, 5, and 4 results for subject A, B, C, and D, respectively.
The number of frames of point cloud data for each result was 540.

\begin{figure}[t]
\centerline{\includegraphics[width=\linewidth]{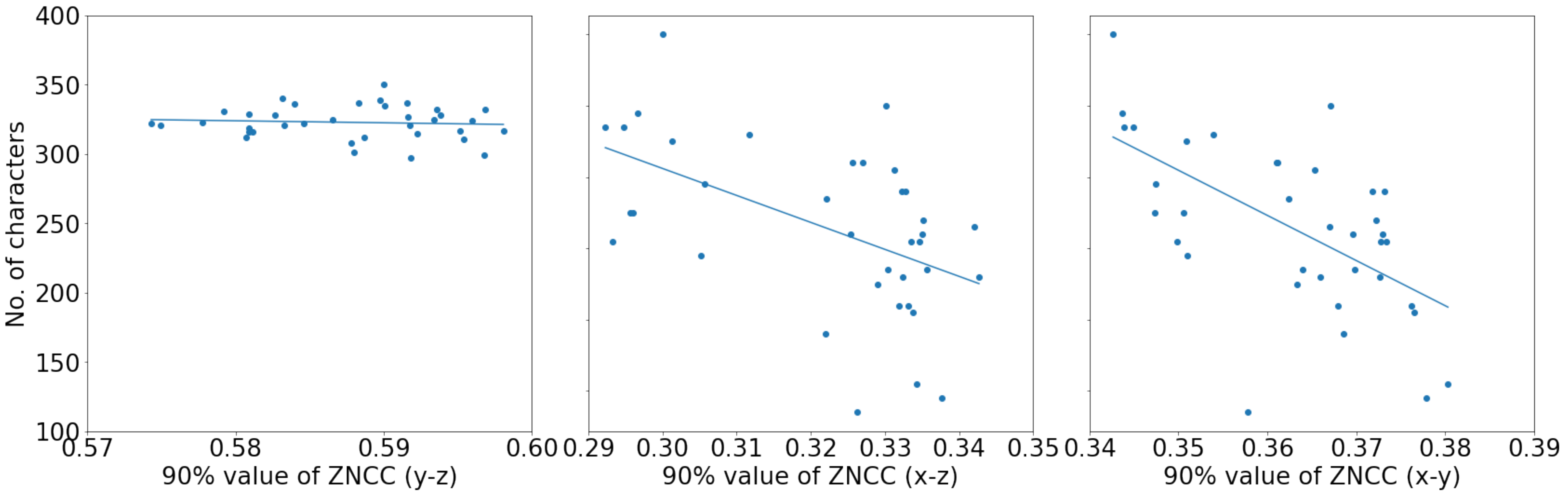}}
\caption{Number of typed characters vs. 90 percentile value of ZNCC for subject A.}
\label{fig_re1}
\end{figure}

\begin{figure}[t]
\centerline{\includegraphics[width=0.8\linewidth]{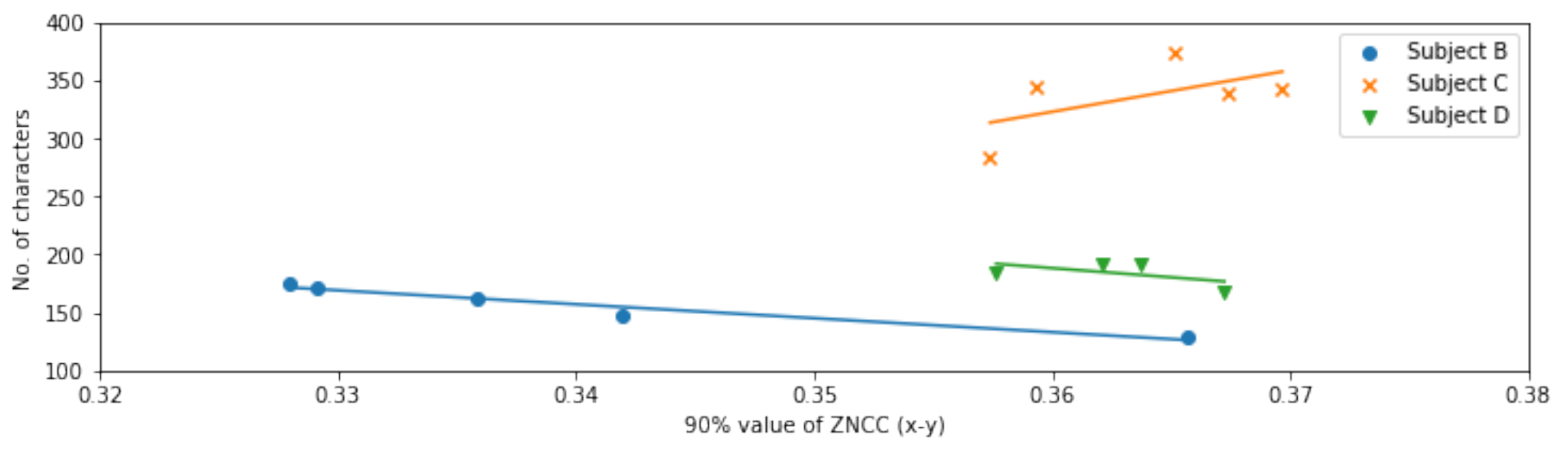}}
\caption{Number of typed characters vs. 90 percentile value of ZNCC for subjects B, C, and D (x-y).}
\label{fig_re2}
\end{figure}

Figure \ref{fig_re1} presents the number of typed characters versus 90 percentile value of ZNCC for subject A. y-z, x-z, and x-y indicate the two-dimensional planes we used for each result. The lines indicate the linear correlation between the horizontal and vertical axes. As seen for the x-z and x-y planes, as ZNCC increases, the number of typed characters increased. There was little correlation for the y-z plane because we removed the information of the x-dimension, which reflects the action of typing well. Figure \ref{fig_re2} presents the number of typed characters versus 90 percentile value of ZNCC for subjects B, C, and D (x-y). We see the same observation for subjects B and D as for subject A. It is counterintuitive that the result for Subject C shows the opposite trend. We need to increase the number of trials for further investigation.

\section{Conclusion}
This paper proposed an edge computing system that enables estimating physical activities of people in offices by using a LIDAR sensor network,
and presented that the proposed system constructs the model for estimating the number of typed characters by ZNCC obtained from point cloud data, through experiments using real LIDAR sensors.
Future work includes an experiment with a larger number of subjects.


\bibliographystyle{unsrt}
{\footnotesize \bibliography{refCCNC2023}}

\end{document}